\documentstyle[aps,preprint,tighten,floats,epsf,rotate]{revtex}

\begin{document}
\draft

%
\title{Resonant Production of Topological Defects}
%
\author{Sanatan Digal, Rajarshi Ray, Supratim Sengupta, and Ajit M. 
Srivastava}
\address{Institute of Physics, Sachivalaya Marg, Bhubaneswar 751005, 
India}
%
%
\maketitle
\widetext
\parshape=1 0.75in 5.5in
\begin{abstract}

 We describe a novel phenomenon in which vortices are produced due to 
resonant oscillations of a scalar field which is driven by a periodically 
varying temperature $T$, with $T$ remaining much below the critical 
temperature $T_c$. Also, in a rapid heating of a localized region to a 
temperature {\it below} $T_c$, far separated vortex and antivortex can 
form. We compare our results with recent models of defect production 
during reheating after inflation. We also discuss possible experimental 
tests of our predictions of topological defect production {\it without} 
ever going through a phase transition.

\end{abstract}
\vskip 0.125 in
\parshape=1 -.75in 5.5in
\narrowtext
 
   Recently, a lot of interest has been focussed on the formation and 
consequences of topological defects both in the context of particle 
physics models of the early universe \cite{vil} as well as in condensed 
matter systems \cite{zurek}. In all the investigations, defects are 
produced by two different mechanisms. They could be produced by external 
influence, such as flux tubes in superconductors in an external magnetic 
field or vortices in superfluid helium in a rotating vessel. The only 
other method of producing topological defects is either during a phase 
transition (via Kibble mechanism \cite{kbl}), or due to thermal 
fluctuations (with defect density 
suppressed by the Boltzmann factor). It has been recently 
demonstrated by some of us \cite{flip} that defects can be produced via 
a new mechanism due to strong oscillations, and subsequent flipping, of 
the order parameter (OP) field during a phase transition. 

 In this letter we discuss a new phenomenon where defects can form without
ever going through a phase transition. In our model, defects form due to 
resonant oscillations of the OP field at a temperature 
which is periodically varying, but remains much below the critical 
temperature $T_c$. We also show that in rapid heating of a localized 
region to a temperature {\it below} $T_c$, far separated vortex and 
antivortex (or a large string loop in 3+1 dimensions) can form. 
Our model Lagrangian describes a system with a spontaneously broken 
global U(1) theory in 2+1 dimensions. The Lagrangian is expressed in 
terms of scaled, dimensionless variables,

\begin{equation}
{\cal L} = \frac{1}{2}(\partial_{\mu}\Phi^{\dagger})(\partial^{\mu}\Phi) 
- \frac{1}{4}(\phi^2-1)^2 - {\alpha \over 8} T^2 \phi^2 .
\end{equation}

 Here $\Phi=\Phi_{1}+i\Phi_{2}$ is a complex scalar field with 
magnitude $\phi$. $T$ is the temperature of the system (with $T_c$ = 1) 
and $\alpha$ is a dimensionless parameter. We take $\alpha=4$. We get 
similar results with linear temperature dependence in the effective 
potential.  We emphasize that the basic physics of our
model resides in the time dependence of the effective potential.
We achieve this by using a time dependent temperature. One could also 
do this by periodically varying some other parameter such as pressure 
(which may be experimentally more feasible), or even a time-dependent 
external electric or magnetic field (say, for liquid crystals 
\cite{expl}). The oscillatory temperature dependence 
in our model leads to field equations which are similar to those with an 
oscillating inflaton field coupled to another scalar field in the models
of post-inflationary reheating \cite{in1,in11,in2,in3,in4} (more 
precisely, to the case with spontaneously broken symmetry for the scalar 
field  \cite{in2}), with the oscillating temperature playing the role 
of the inflaton field. Analysis of growth of fluctuations 
shows the existence of exponentially growing modes in this case implying 
that fluctuations will grow rapidly for modes with wave vector lying in 
the resonance band.  However, there are also crucial differences between 
our results and those in the literature \cite{in1,in11,in2,in3,in4}, as we 
will explain later.

 We use the following equations for field evolution.

\begin{equation}
\partial^2 \Phi_i/\partial t^2 + \eta \partial\Phi_i/\partial t - 
\bigtriangledown^2 \Phi_i + V^\prime(\phi) = 0 .
\end{equation}
 
 Here $V(\phi)$ is the effective potential in Eqn.(1) and $\eta$ is a 
dissipation coefficient. We solve Eqn.(2) using a second order staggered 
leapfrog algorithm. We do not include a noise 
term in Eqn.(2). The basic physics we discuss does 
not depend on noise. Also, as discussed above, time dependence of 
the effective potential could be thought to arise from some other source 
than the temperature (with temperature kept low to suppress any 
thermal fluctuations). Here we mention that the effect of noise on the 
growth of fluctuations has been studied in ref. \onlinecite{in5} in the 
context of reheating after inflation. It is shown in ref.\onlinecite{in5}
that small amplitude noise does not affect growth of fluctuations for
unstable modes. In a future work we will study the effect 
of noise in our model, and also explore connections with the well 
studied phenomenon of {\it stochastic resonance} in condensed matter 
systems \cite{stch}.

  We first discuss a situation in which a central patch in the system 
is instantaneously heated from $T = 0$ to $T = T_0 < 
T_{c}$ and maintained at that temperature, while the region outside the 
patch is kept at $T$ = 0. Field is evolved by Eqn.(2), with $T$ being 
the local temperature. The temperature profile of the region 
separating the central patch from the surrounding region is taken to be 
$T(r) = (T_0/2) (1 - tanh((r-R)/\bigtriangleup))$; R is the radius of the
(circular) central patch and $\bigtriangleup$ is the wall thickness. 
(Our results do not depend on the profile used for this boundary region.) 
We use a 500 $\times $500 lattice with physical size equal to $80 
\times 80$. Time step $\Delta t$ was taken to be 0.008. We take $R = 
24$ and $\bigtriangleup = 0.32$. Initial magnitude of $\Phi$ was taken 
to be equal to the $T = 0$ vacuum expectation value (vev), 
while its phase varied linearly from 0 to $2\pi$ along the Y-axis, being 
uniform along the X-axis. (This choice facilitated the use of periodic 
boundary conditions to evolve the field. Defect production happens with
much smaller phase variation also.) 
The instantaneous heating of the central region with $T_0 = 0.72$ (with
$\eta$ = 0.4), led to destabilization of $\Phi$ in that region. 
The field, while trying to relax to its new equilibrium value, overshot 
the central barrier and ended up on the opposite side of the vacuum 
manifold. As discussed in \cite{flip}, this results in the formation of a 
vortex at one boundary of the {\it flipped} region, with an anti-vortex 
forming at the opposite boundary, as shown in Fig.1. This occurs because 
overshooting of $\Phi$ (in a localized region) leads to a 
discontinuous change in its phase by $\pi$ ({\it flipping} of $\Phi$). 
Note that these defects could not have formed via the Kibble mechanism 
as the system never goes through a symmetry breaking phase transition
in our case. Similarly, thermal fluctuations can not directly produce a 
pair in which defect and antidefect are so far separated.
 
  This shows that above scenario can be used to provide
an unambiguous experimental verification of the flipping mechanism 
described in \cite{flip} for defect production. For example, one can 
locally heat a central portion of superfluid $^4$He system to a 
temperature $T_0<T_c$ ($T_0$ should be large enough to allow for 
overshooting \cite{dcc}). In $3+1$ dimension, this would lead
to the formation of a string loop, with a size of order of the size of 
the heated central region (in $2+1$ dimensions one will get far separated 
vortex-antivortex pair). The requirement of small phase variation 
can be easily achieved for superfluid helium by allowing for small, 
uniform,  superflow across the system. (For superconductors, one can 
allow small supercurrent.) The fact that the formation of such a large 
string loop does not require a phase transition, makes this a clean 
signal, unpolluted by the presence of smaller string loops. 
 
 We now describe the resonant production of 
vortex-antivortex pairs, induced by periodically varying temperature $T$. 
$T$ (in Eqn.(1)) is taken to be spatially uniform, with its periodic 
variation given by $T(t) = T_{0} + T_{a}\sin(\omega t)$. The choice of 
frequency $\omega$ was guided by the range of frequency 
required to induce resonance for the case of spatially uniform field
(evolved by Eqn.(2)). We find that resonance happens when 
$\omega$ lies in a certain range. (We are assuming
that for the relevant range of $\omega$  here, the system can be
considered in quasi-equilibrium so that the use of temperature dependent
effective potential makes sense.) This frequency range for which resonant 
defect production occurs depends on the average temperature ($T_{0}$) as 
well as on the amplitude of oscillation $T_a$, with the range becoming 
larger as $T_{0}$ approaches $T_c$. 

 Initially the magnitude of $\Phi$ is taken to be equal to the vev at 
$T = T_0$, over the entire lattice. The phase $\theta$ of $\Phi$ was 
chosen to have small random variations at each lattice point, with
$\theta$ lying between $20^{0}$ and $40^{0}$.  We have carried out 
simulations for other widely different initial configurations as well. 
In one case we took the domain structure 
developing at late stages in above simulations, 
and used it as the initial data for evolution with a different set of 
parameters. In another case, we took the lattice to consist of only 
four domains with small variations of $\theta$ between domains. 
Defects are always produced as long as there is some {\it non-uniformity} 
in the initial phase distribution. Although we find that at early 
times defect formation depends on the initial configuration, the {\it 
asymptotic average defect density} seems to remain unchanged. 
Note that the analysis based on the Mathieu 
equation, \cite{in1,in11,in2,in3,in4} in the context of inflationary 
models, suggests that the growth of fluctuations may depend on the initial 
configuration. However, such analysis cannot 
be trusted for times beyond which fluctuations become too large. 

 Fig.2 shows a portion of the lattice for $\omega = 
1.19$ and $\eta=0.013$, with four 
vortex-antivortex pairs. We count only those pairs which are separated
by a distance larger than 2$m_H^{-1}$. We emphasize again, formation of
these vortices cannot be accounted for by Kibble mechanism, or by the 
thermal production of defects. To compare our results
with the  defect density expected in the case of thermally produced 
defects, we have estimated the energy of a defect-antidefect pair, using 
the numerical techniques in ref.\onlinecite{emin}. We find it to be 
about 2.5 (for $T_0 = 0.31$) when the separation between the vortex and 
the antivortex is 2$m_H^{-1}$ (with temperature dependent $m_H$). If we 
take thermally produced defect density to be of order 
$\sim T^2$ exp(-$E_{pair}/T$), then even with $T = T_0 + T_a $ = 0.435, 
the defect density is only about $6 \times 10^{-4}$, which is less than
the defect densities we find.

 We now discuss differences between defect production in our model and
those discussed in refs. \onlinecite{in2,in3,in4} in the context of 
reheating after inflation. In these works,
non-thermal fluctuations grow large enough to {\it restore} the 
symmetry \cite{in11}. Topological defects are produced when the 
symmetry subsequently {\it breaks} due to rescattering effects and/or 
universe expansion.  However, in our case there is {\it no} 
symmetry restoration ever (see, also, ref.\onlinecite{in6}). 
To check this we plot the probability 
distribution of $\Phi$ in Fig.3. This clearly shows that
the peak of the probability distribution lies around a valley of 
radius $\simeq$ 0.9 (for $T_0 = 0.375$) and {\it not} at the zero of 
$\Phi$ as would have been the case if symmetry had been restored
\cite{in3}. In all the cases we have studied, we never find the peak 
of the probability density of $\Phi$ to be centered near $\Phi = 0$ 
(except for early transient stages when $\Phi$  in the entire lattice
flips through $\Phi = 0$).
Further, detailed field plots show that the defect production in our 
model happens via the flipping mechanism and not by the formation of 
domain structure (via Kibble mechanism) as would be expected if 
defects were produced due to non-thermal symmetry restoration and its 
subsequent breaking.  Hence the defects are always produced 
in {\it pairs} in our case. Also, the pair production continues 
for all times because of continued {\it localized} flipping of $\Phi$ 
under the influence of the periodically oscillating temperature
(equivalently, an oscillating inflaton field). 

 We have carried out simulations for various values of $T_{0}$, (with 
$\omega=1.19$, $T_{a} = 0.125$, and $\eta = 0.013$ kept fixed), starting
with similar initial configurations. As a function of time, defect 
density rises from zero, and eventually fluctuates about an average value.
We carry out the evolution till $t \sim 1000$ when this average defect 
density becomes reasonably constant. For this set of parameters, 
the smallest value of  $T_{0}$, for which we observe resonant defect 
production, was equal to 0.31. (For linear temperature dependence in
$V(\phi)$ in Eqn.(1) we get defect production at much lower
temperatures, with smallest value of $T_0 = 0.10$ and $T_a$ = 0.08, 
$\omega$  = 1.19.) Fig.4 shows temporal variation of defect 
density at large times.  The fluctuations in defect density are due to 
pair annihilations and due to pair creation of vortices which keeps 
happening periodically because of localized flipping of $\Phi$. 

 Fig.5 shows plot of the asymptotic average defect density (for
$900 \leq t \leq 1000$) as a function of  $T_{0}$. One can 
see that defect density peaks at a value of $T_0 \simeq 0.5$.  This
drop in average defect density for larger temperatures can be
due to decrease in $m_H$. Due to flatter $V_{eff}$ near the true
vacuum, $\Phi$ does not gain too much potential energy when $V_{eff}$
changes due to change in temperature (with $T_a$ = 1.0), which may make
flipping of $\Phi$ difficult (even though the barrier height is smaller
now). A near perfect Gaussian fit to the values of defect density in 
Fig.5 may be indicative of a possibility of establishing some sort of 
correspondence with a thermodynamic phase of the system with some {\it 
different} effective temperature. Investigation of this possibility 
requires accurate determination of asymptotic defect density and its
dependence on different parameters, such as $T_0$, $T_a$, and $\omega$.
(As one can see from Fig.4, average defect densities show slow variation
over a large time scale for certain cases.)  

 We find that there is an optimum value of frequency $\omega$ for which 
the defect density is maximum. It is reasonable to expect that if 
$\omega$ is too large, it would be difficult for the field to overshoot 
the central barrier since larger frequency will lead to averaging
out of the force on $\Phi$ due to the rapidly changing 
effective potential. On the other hand, if $\omega$ is too 
small, the field may have enough time to relax to its equilibrium value 
without being sufficiently destabilized, making the overshooting 
of $\Phi$ more difficult. By keeping all other parameters 
fixed, ($T_{0}=0.5$, $T_{a}=0.13$ and $\eta=0.013$) we have explored 
the behavior of defect density for different values of frequency. 
(For the above parameters, resonant defect production happens for
$0.94 < \omega < 1.31$). For $\omega$ = 0.94, 1.0, 1.13, 1.19, and
1.31, we find the average defect density to be 0.0048, 0.0061,
0.0059, 0.0058, and 0.0042 respectively.

 We have made some check on the dependence of the asymptotic average 
defect density on $T_{a}$. For $T_{0}=0.5$, $\omega=1.19$ and 
$\eta=0.013$, the average defect density is larger for $T_a = 0.13$ 
compared to the case when $T_{a}=0.25$. (Again, note that we are
focusing on asymptotic defect density.) When we increase $\eta$, it 
results in suppression of the defect density since damping makes it 
difficult for $\Phi$ to overshoot the barrier. Moreover, as $\eta$ 
is gradually increased, the minimum value of $T_{0}$ required to 
induce resonant oscillations also increases and for $\eta > 0.63$, 
resonant production of vortices is not possible for $T_{0} < T_{c}$. We 
also find that the range of frequency for which resonant defect 
production occurs, becomes narrower on increasing $\eta$.   
  
 In conclusion, we have demonstrated an interesting phenomenon where 
topological defects
can form at very low temperatures, without the system ever going through 
the phase transition. It should be easy to experimentally verify this
scenario of defect production in many condensed matter systems, such
as superfluid helium, or superconductors. There are also important 
implications of our results for defect production during reheating after 
inflation. Earlier studies \cite{in11,in2,in3,in4} have focussed on defect 
production due to non-thermal symmetry restoration and its subsequent 
breaking. Our results show the possibility that defects may be produced 
due to oscillations of the inflaton field via the flipping mechanism even 
when there is {\it no} symmetry restoration ever. This may lead to defect 
production under more generic conditions than discussed in the literature. 
Our results suggest the very interesting possibility that various  models 
of re-heating after inflation can have analogs in the condensed matter 
systems which may provide possible experimental tests for the predictions
of these models.


\newpage
\centerline {\bf FIGURE CAPTIONS}
\vskip .1in

1) The central region is instantaneously heated to a temperature 
$T_0$ = 0.72.  A far separated vortex-antivortex pair 
is produced due to flipping of $\Phi$ in the heated region.

2) $\Phi$ at t = 908.0 for the resonant
oscillation case for a portion of lattice, showing randomly oriented 
domains, with 4 vortex-antivortex pairs. $T_0$ = 0.31 and $T_a$ = 
0.125 for this case.

3) (a) Probability density of $\Phi$ at $t$ = 328.0 
(b) Corresponding contour plot. Darker portions signify larger 
probability.

4) Evolution of defect density $n$ at large times.
Straight lines show linear fits to the respective
curves. $n$ for $\omega$ = 1.19 and 1.31, are shown by 
the solid and the dotted curves respectively (with $T_0 = 0.5$ and 
$T_a$ = 0.125). Dashed curve shows $n$ for 
$T_0$ = 0.31, $T_a$ = 0.125 and $\omega$ = 1.19.

5) Variation of asymptotic average defect density with $T_0$.
Dots show the values obtained from numerical simulations, while the
curve shows Gaussian fit to these points.

\newpage

\vskip -0.25in
\begin{figure}[h]
\begin{center}
\leavevmode
\epsfysize=12truecm \vbox{\epsfbox{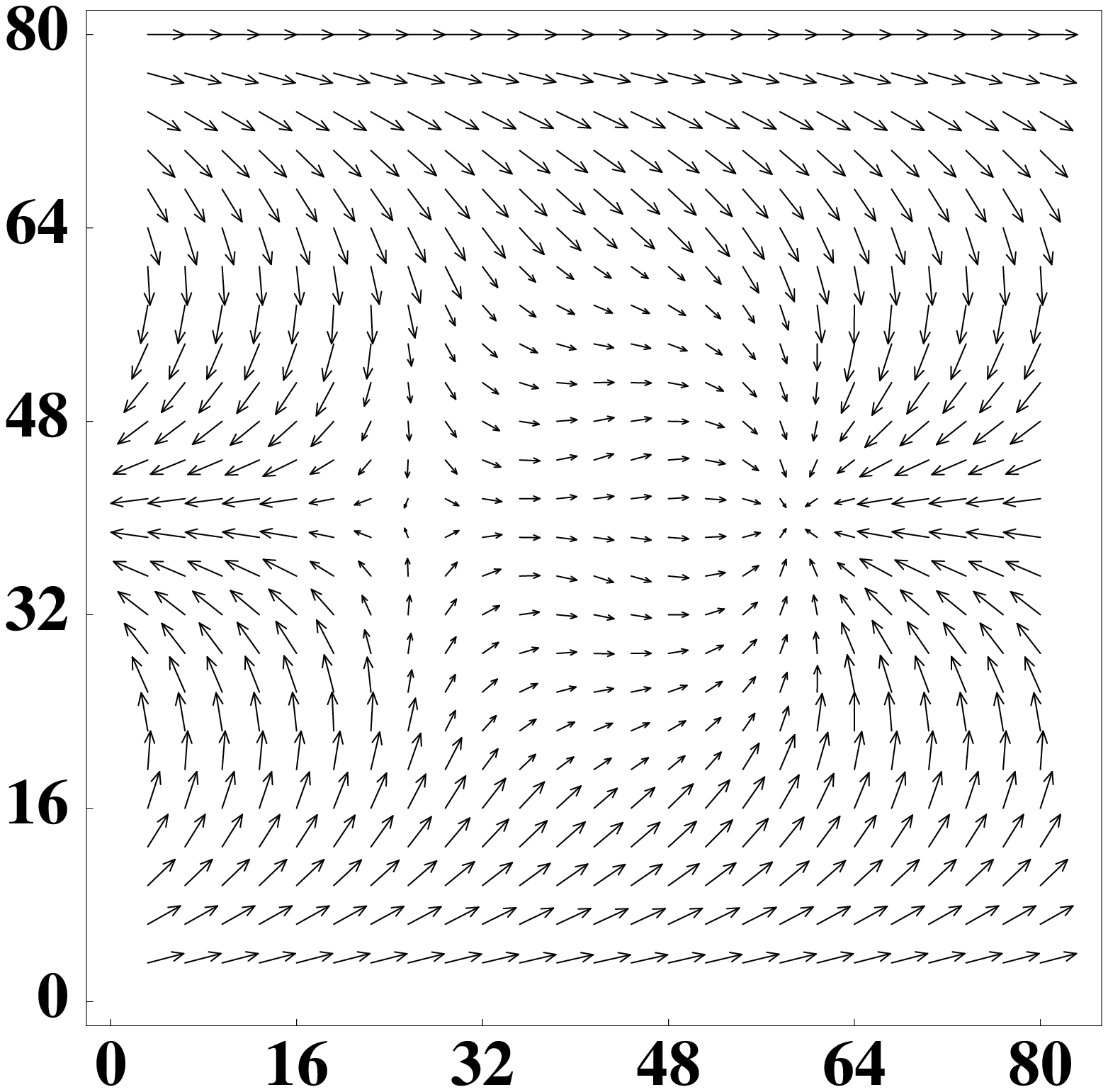}}
\end{center}
\vskip -1in
\caption{}
\label{Fig.1}
\end{figure}

\newpage

\vskip -0.35in
\begin{figure}[h]
\begin{center}
\leavevmode
\epsfysize=15truecm \vbox{\epsfbox{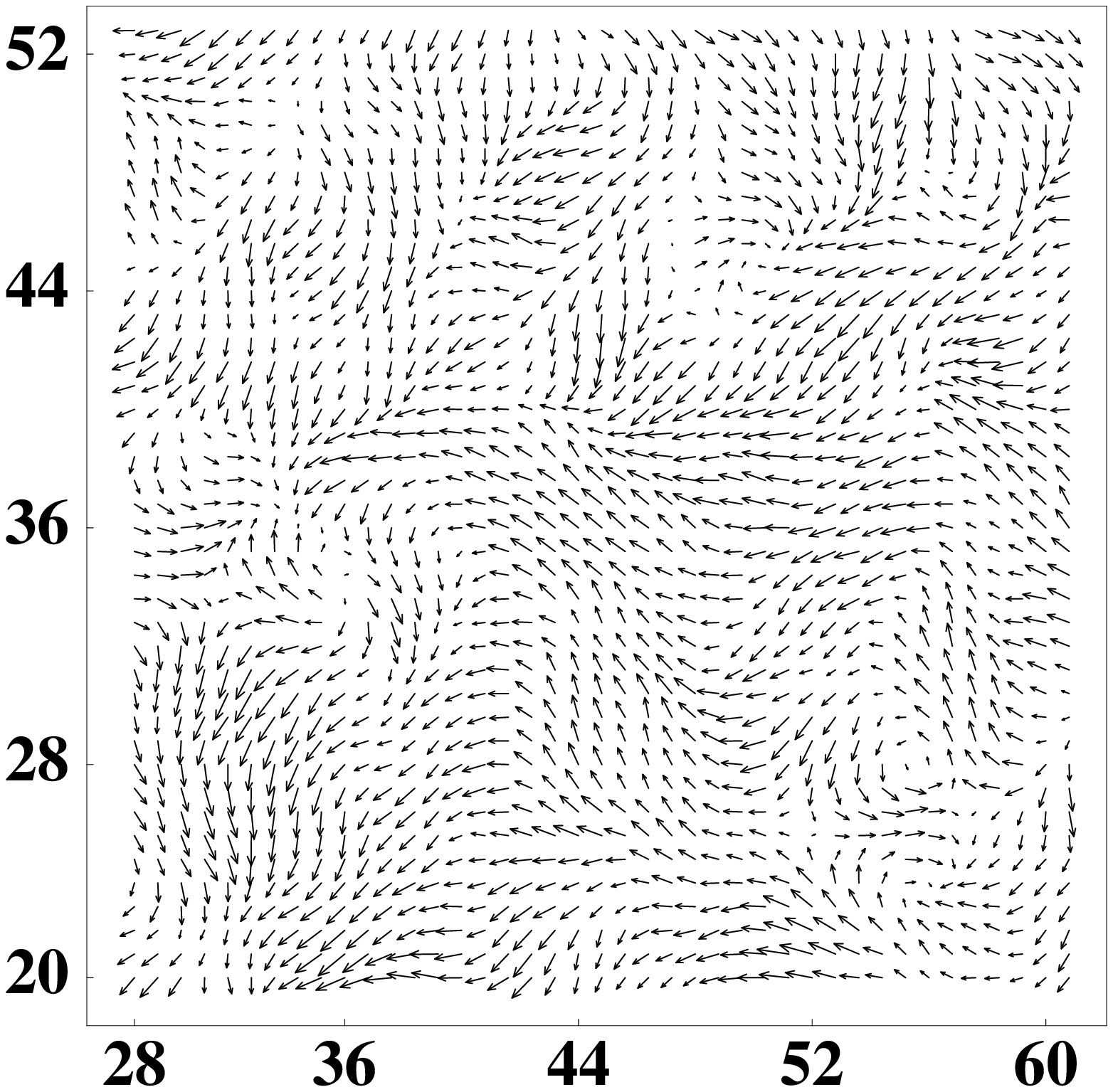}}
\end{center}
\vskip -1.5in
\caption{}
\label{Fig.2}
\end{figure}

\newpage

\begin{figure}[h]
\begin{center}
\leavevmode
\vskip -1.85in
\epsfysize=20truecm \vbox{\epsfbox{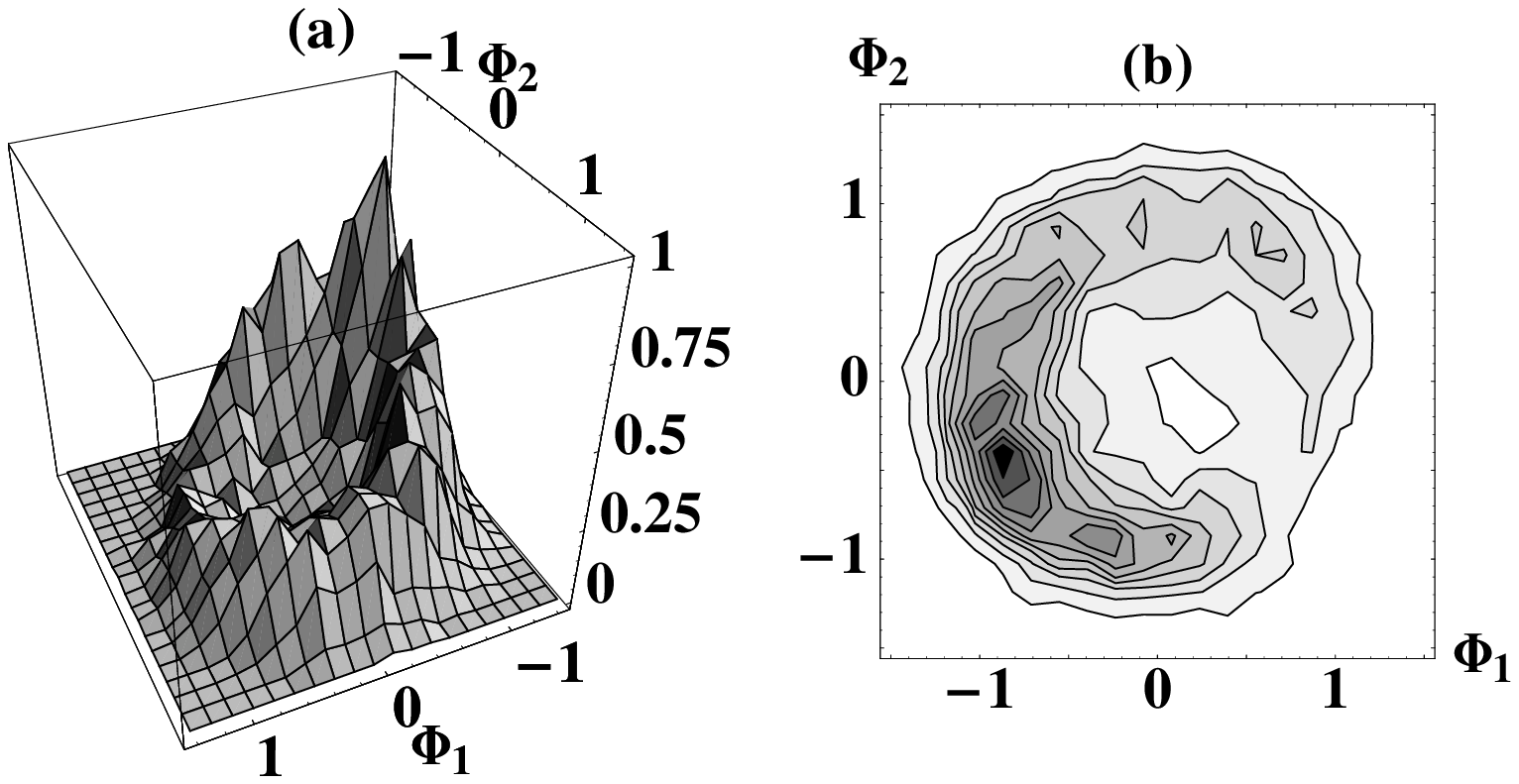}}
\end{center}
\vskip -2in
\caption{}
\label{Fig.3}
\end{figure}

\newpage

\begin{figure}[h]
\begin{center}
\leavevmode
\vskip -0.75in
\epsfysize=20truecm \vbox{\epsfbox{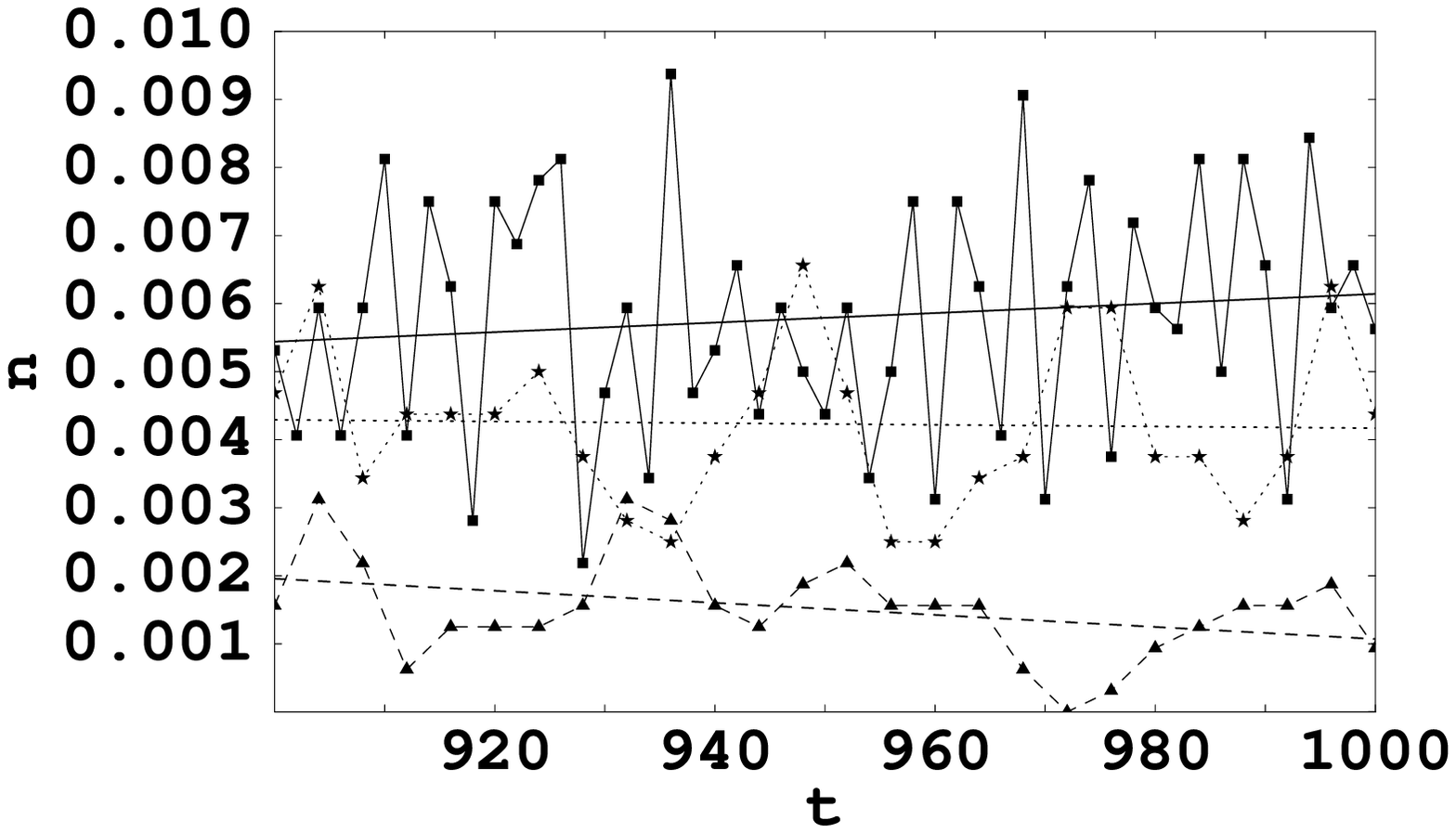}}
\end{center}
\vskip -2.0in
\caption{}
\label{Fig.4}
\end{figure}

\newpage

\begin{figure}[h]
\begin{center}
\leavevmode
\vskip -1.25in
\epsfysize=20truecm \vbox{\epsfbox{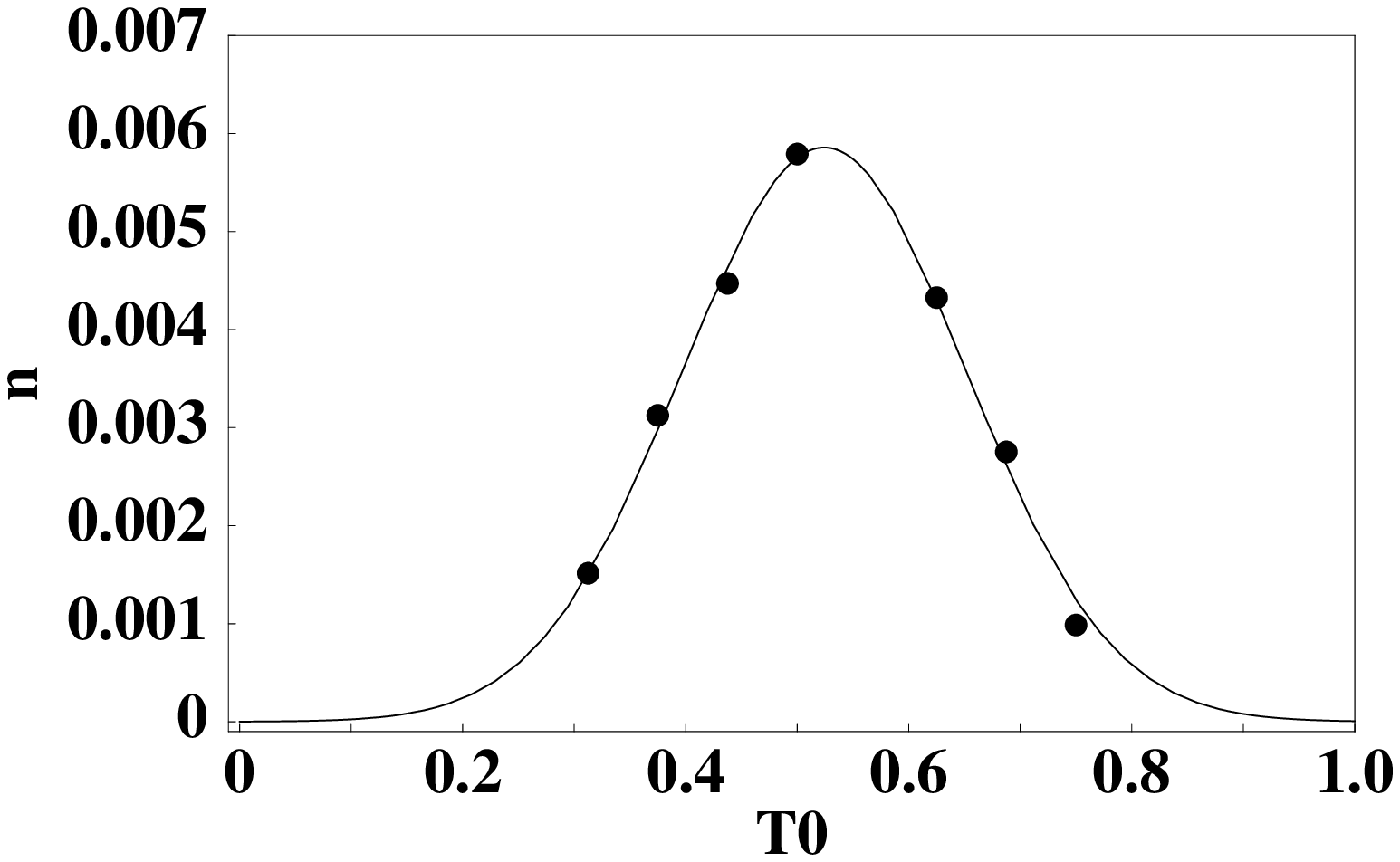}}
\end{center}
\vskip -2in
\caption{}
\label{Fig.5}
\end{figure}


\begin{thebibliography}{99}

\bibitem{vil} A.Vilenkin and E.P.S.Shellard, ``Cosmic 
Strings and Other Topological Defects", (Cambridge University Press, 
Cambridge, 1994).

\bibitem{zurek} W.H. Zurek, Phys. Rep. {\bf 276},
177 (1996). 

\bibitem{kbl} T.W.B. Kibble, J. Phys. {\bf A9}, 1387 (1976).

\bibitem{flip} S. Digal, S. Sengupta, and A.M. Srivastava,
Phys. Rev. {\bf D55}, 3824 (1997); {\it ibid} Phys. Rev. {\bf D56},
2035(1997).

\bibitem{expl} S. Digal, S. Sengupta, and A.M. Srivastava,
Phys. Rev. {\bf D58}, 103510 (1998).

\bibitem{in1} J.H. Traschen and R.H. Brandenberger, Phys. Rev.
{\bf D42}, 2491 (1990); Y. Shtanov, J. Traschen, and R. Brandenberger,
Phys. Rev. {\bf D51}, 5438 (1995); L. Kofman, A. Linde, and A.A. 
Starobinsky, Phys. Rev.  {\bf D56}, 3258 (1997).

\bibitem{in11} I.I. Tkachev, Phys. Lett. {\bf B376}, 35 (1996); L. 
Kofman, A. Linde, and A.A. Starobinsky, Phys. Rev. Lett. {\bf 76}, 
1011 (1996); S. Khlebnikov, L. Kofman, A. Linde, and I. Tkachev,
Phys. Rev. Lett. {\bf 81}, 2012 (1998).

\bibitem{in2} M.F. Parry and A.T. Sornborger, Phys. Rev. {\bf D60}, 
103504 (1999).

\bibitem{in3} I.Tkachev, S. Khlebnikov, L. Kofman, and A. Linde,
Phys. Lett. {\bf B440}, 262 (1998).

\bibitem{in4} S. Kasuya and M. Kawasaki, Phys. Rev. {\bf D58}, 083516 
(1998)

\bibitem{in5} V. Zanchin, A. Maia, Jr., W. Craig, and R. Brandenberger,
Phys. Rev. {\bf D60}, 023505 (1999).

\bibitem{stch} L. Gammaitoni, et.al. ``Stochastic Resonance", Rev. Mod.
Phys. {\bf 70}, 223 (1998)

\bibitem{dcc} S. Digal, R. Ray, S. Sengupta, and A.M. Srivastava, 
hep-ph/9805227. 

\bibitem{emin} A.M. Srivastava, Phys. Rev. {\bf D 47}, 1324 (1993).

\bibitem{in6} D. Boyanovsky, H.J. de Vega, R. Holman, and
J.F.J. Salgado, Phys. Rev. {\bf D54}, 7570 (1996).

\end{thebibliography}
\end{document}